\documentstyle[epsfig,12pt]{article}
\begin{document}


\def\etal{{\it et.~al.}}
\def\ie{{\it i.e.}}
\def\eg{{\it e.g.}}

\def\be{\begin{equation}}
\def\ee{\end{equation}}
\def\bea{\begin{eqnarray}}
\def\eea{\end{eqnarray}}
\def\bean{\begin{eqnarray*}}
\def\eean{\end{eqnarray*}}
\def\bary{\begin{array}}
\def\eary{\end{array}}
\def\bi{\bibitem}
\def\bit{\begin{itemize}}
\def\eit{\end{itemize}}

\def\lan{\langle}
\def\ran{\rangle}
\def\lra{\leftrightarrow}
\def\la{\leftarrow}
\def\ra{\rightarrow}
\def\dash{\mbox{-}}

\def\re{\rm Re}
\def\im{\rm Im}
\def\eps{\epsilon}
\def\sg{\tilde g}
\def\sb{\tilde b}
\def\st{\tilde t}
\def\vk{\bf k}
\def\vp{\bf p}
\def\ua{\uparrow}
\def\da{\downarrow}


\Large \centerline {\bf $Z$ DECAYS INTO LIGHT GLUINOS:}
\centerline{\bf a calculation based on unitarity \footnote{Enrico
Fermi Institute preprint EFI 03-02, to appear in Phys. Rev. D.}}
\normalsize

\vskip 2.0cm

\centerline {Zumin Luo~\footnote{zuminluo@midway.uchicago.edu}}
\vskip 0.5cm

\centerline {\it Enrico Fermi Institute and Department of Physics}
\centerline{\it University of Chicago, 5640 S. Ellis Avenue,
Chicago, IL 60637}

\vskip 4.0cm

\begin{quote}
The $Z$ boson can decay to a pair of light gluinos through
loop-mediated processes.  Based on unitarity of the $S$-matrix,
the imaginary part of the decay amplitude is computed in the
presence of a light bottom squark. This imaginary part can provide
useful information on the full amplitude. Implications are
discussed for a recently proposed light gluino and light bottom
squark scenario.

\end{quote}

\bigskip

\noindent

PACS Categories: 11.30.Pb, 12.60.Jv, 13.38.Dg, 14.80.Ly

\vfill
\newpage

\section{Introduction \label{sec:int}}

A relatively light (12--16 GeV) gluino $\sg$, along with a lighter
(2--5.5 GeV) bottom squark $\sb$, has been proposed
\cite{Berger:2000mp} to explain the excess of the cross section
for bottom quark production at hadron colliders. The $\sb$ squark
is assumed to be a mixture of $\sb_L$ and $\sb_R$, the
superparterners of $b_L$ and $b_R$. Other supersymmetric (SUSY)
particles, except the other bottom squark $\sb'$ and one of the
top squarks, are assumed to be sufficiently heavy. The masses of
$\sb'$ and the light top squark $\st$ are constrained by the
electroweak data to be below 180 GeV and 98 GeV, respectively
\cite{Cho:2002mt}. We follow the convention in
Ref.~\cite{Berger:2000mp} to define
\be \left( \begin{array}{c} \sb \\ \sb' \end{array} \right) =
\left(
\begin{array}{c c} \cos \theta_{\sb} & \sin \theta_{\sb} \cr
                        - \sin \theta_{\sb} & \cos \theta_{\sb}
\end{array} \right)
\left( \begin{array}{c} \tilde b_R \\ \tilde b_L \end{array}
\right)~~~.
\ee
The introduction of these new particles gives rise to new
interactions in various processes. For example, the total decay
width of the $\Upsilon$ is raised since the decay $\Upsilon \to
\sb\sb^*$ \cite{Berger:2001jb} is now permitted; the decay width
of the $Z$ boson is also changed \cite{Clavelli:1996zm,
Cao:2001rz}. As a result, the extraction of the strong coupling
constant $\alpha_s$ at these two mass scales will be affected. By
contributing to the $\beta$-function, these SUSY particles slow
down the evolution of $\alpha_s$ with energy scale
\cite{Clavelli:zv}. The situation has recently been studied in
detail by Chiang {\it et al.} \cite{Chiang:2002wi} and no
clear-cut decision can be made in favor of either the Standard
Model evolution or the evolution in the light gluino/light bottom
squark scenario. The partial decay width $\Gamma(Z \to \sg \sg)$
remains a key quantity to be determined. A better evaluation of
$\Gamma(Z \to \sg \sg)$, among other things, can improve our
understanding of the effect of these new particles on the
electroweak measurables at the $Z$ pole and hence the
determination of $\alpha_s(M_Z)$ in the scenario.

To validate the proposition of these new particles, direct
searches for light gluinos and light bottom squarks at $e^-e^+$
colliders will definitely play a key role. An analysis has been
presented recently by Berge and Klasen \cite{Berge:2002ev} of
gluino pair production at linear $e^-e^+$ colliders. However, they
only considered the mass range $m_{\sg} \ge 200$ GeV. Production
of light gluino pairs was studied by Ref.~\cite{Nelson:1982cu} and
its updated version \cite{Kileng:1994vc}. However, a
chirally-mixed {\it light} bottom squark was not included in
either of these calculations. The decay of on-shell $Z$ bosons
into gluino pairs was first discussed in Refs.~\cite{Campbell:kf,
Kane:xp}, but no chiral mixing between squarks was allowed.
Djouadi and Drees \cite{Djouadi:1994js} took into account chiral
mixing and computed an explicit expression for $\Gamma(Z \to \sg
\sg)$. However, they neglected the gluino mass and required all
squark mass eigenstates to be heavier than $M_Z/2$. Production of
light gluinos at $p{\bar p}$ colliders was considered by Terekhov
and Clavelli \cite{Terekhov:1996rp} but without inclusion of the
light bottom squark either. Therefore, an analysis of light gluino
production in the presence of a light bottom squark will be very
useful for gluino searches.

\begin{figure}
\centerline{\epsfysize = 3.5 in
\epsffile{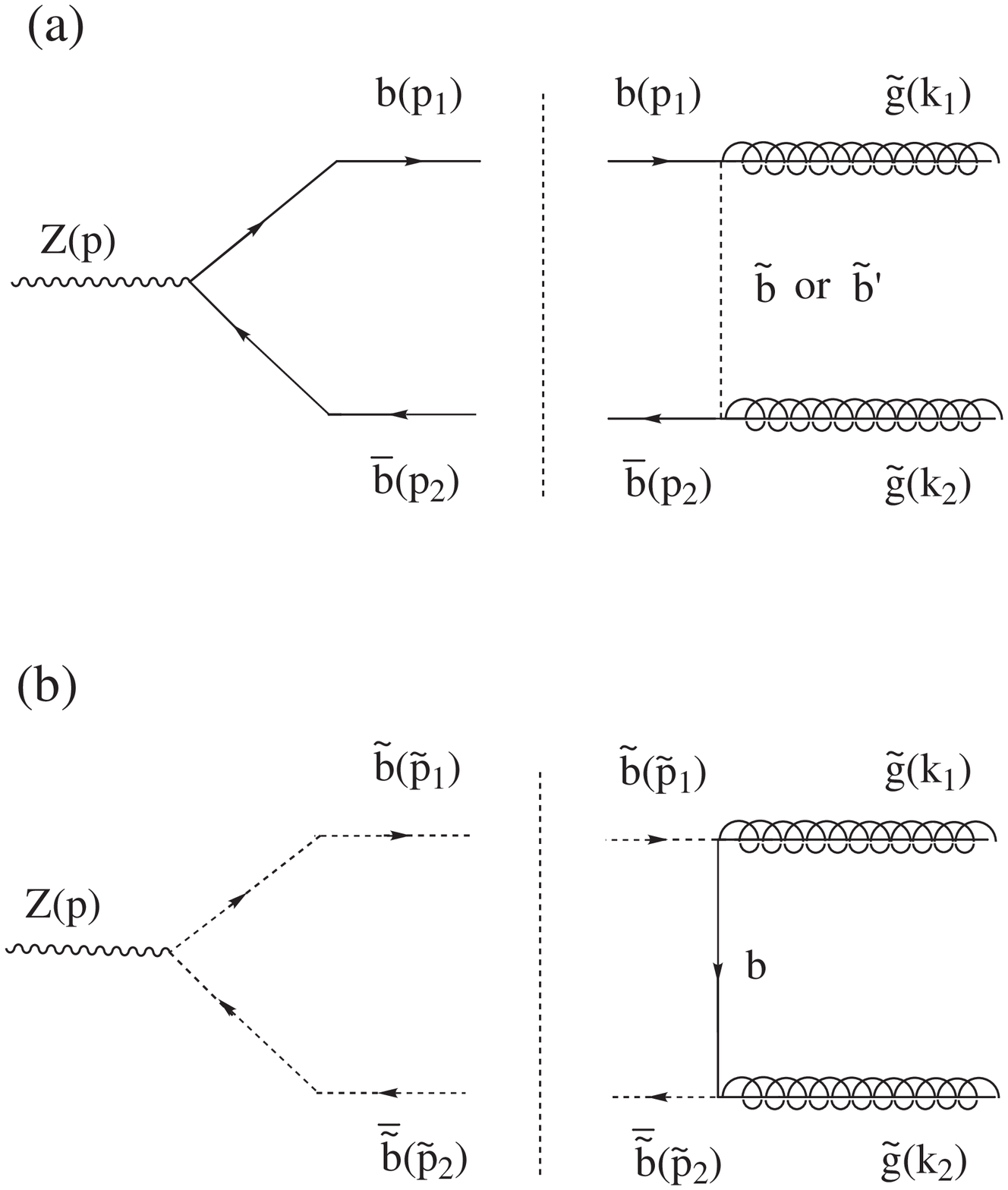}
}
\caption[]{\label{fig:cut}
  Cut Feynman diagrams for $Z \to \sg \sg$: (a) $Z \to (b{\bar b})^*
  \to \sg \sg$, (b) $Z \to (\sb \bar {\sb})^* \to \sg \sg$. Similar
  diagrams with $\sg(k_1) \leftrightarrow \sg(k_2)$ are not shown but
  should be included in the calculation with an overall minus sign. }
\end{figure}

Although all the previous calculations agree with each other that
the branching ratio of $Z \to \sg \sg$ is less than ${\cal
O}(10^{-3})$, they differ in some important features of the
process. At the one-loop level the decay occurs through two types
of diagrams; see Fig.~\ref{fig:cut}. In type (a) diagrams the $Z$
couples to a pair of quarks and a squark is exchanged during the
process, while in type (b) diagrams the $Z$ couples to two squarks
and a quark is exchanged. Refs.~\cite{Berge:2002ev, Campbell:kf}
disagree with Refs.~\cite{Kileng:1994vc, Djouadi:1994js} in the
relative sign between the two types of diagrams. Considering only
non-mixed chiral squarks of equal mass, Kane and Rolnick
\cite{Kane:xp} claimed that the amplitude of the process is
identically zero when $m_q = m_{\tilde q}$ is satisfied for each
supersymmetric pair, even if weak isospin is broken so that, for
example, $m_t \ne m_b$. However, other references
\cite{Berge:2002ev, Kileng:1994vc, Djouadi:1994js} state that for
the contribution of quarks and squarks of a given generation to
vanish, we must have both mass degeneracy in the quark isospin
doublet (e.g., $m_d = m_u$) and mass degeneracy in the squark
isospin doublets (e.g., $m_{\tilde d} = m_{\tilde d'} = m_{\tilde
u} = m_{\tilde u'}$). There are also two contradictory opinions
with regard to cancellation of ultraviolet divergences.
Refs.~\cite{Berge:2002ev, Campbell:kf, Kane:xp} asserted that
ultraviolet singularities cancel separately for each weak isospin
partner, while Djouadi and Drees \cite{Djouadi:1994js} found that
the amplitude is finite only after summing over a complete
isodoublet. This discrepancy is essentially related to the
relative sign between diagrams (a) and (b) in Fig.~\ref{fig:cut}.
The divergent parts of the two diagrams must have opposite signs
for them to cancel separately for each isospin partner. There may
not be any constraint on the relative sign between (a) and (b) for
divergences to cancel within an isodoublet, since the two members
in an isodoublet have opposite $I_3$ (the third component of the
weak isospin) and the divergences are generally proportional to
$I_3$.

A full calculation of $\Gamma(Z \to \sg \sg)$ involves evaluation
of the Feynman diagrams in Fig.~\ref{fig:cut}, with the cut
(s)quark lines connected. To get a meaningful result, one has to
deal with difficult one-loop integrals and remove singularities
due to on-shell particles. In this paper we try to provide a
different approach to solving the above-mentioned discrepancies.
Since $2 m_b < M_Z$ and $2 m_{\sb} < M_Z$ in the proposed
scenario, the decay amplitude has an imaginary part which is
finite and can be calculated in an easier way. It is likely that
the imaginary part can provide some useful information on the full
amplitude. Similar situations arise in the $K_S$--$K_L$ mass
difference and the decay $K_L \to \mu^+ \mu^-$ \cite{Sehgal:db}.
In each case the high-momentum components of the loop diagrams are
suppressed (through the presence of the charmed quark
\cite{Gaillard:1974hs}), leaving the low-mass on-shell states
($\pi \pi$ or $\gamma \gamma$, respectively) to provide a good
estimate of the matrix element. In the light gluino and light
bottom squark scenario, the decay width for $Z \to \sg \sg$
usually turns out to be only a few times larger than the
contribution from the imaginary part alone.

This paper is organized as follows. Section~\ref{sec:unit}
establishes the unitarity relation of the ${\cal M}$ matrix
elements. Amplitudes of the cut diagrams are calculated in
Section~\ref{sec:amps} and the results are listed in the Appendix.
The lower bound based on the imaginary part of the decay amplitude
for $Z \to \sg \sg$ is presented in Section~\ref{sec:lb}.
Implications of the imaginary part for the full amplitude are
discussed in Sections~\ref{sec:fdw}. Implications for gluino
searches and running of $\alpha_s$ are discussed in Sections
\ref{sec:gs} and \ref{sec:ra}, respectively. Section~\ref{sec:sum}
summarizes.

\section{Unitarity relation
\label{sec:unit}}

Let us first review the decays $K_{L,S} \to l^- l^+$ considered in
Ref.~\cite{Sehgal:db}. As is the case with $Z \to \sg \sg$, both
decays are forbidden at the tree level. However, they can occur
through a two-photon ($\gamma\gamma$) intermediate state. Other
intermediate states such as $\pi\pi\gamma$ and $3\pi$ are much
less important. As a consequence of the unitarity of the
$S$-matrix ($S^{\dagger}S\equiv (1+ i T)^{\dagger}(1+i T)=1$), the
$T$-matrix element between the initial state $K_{L,S}$ and the
final state $l^- l^+$ satisfies the following relation
\be {\rm Im}\left [\langle l^-l^+|T|K_{L,S}\rangle \right ] =
\frac{1}{2}\left [\langle l^-l^+|T^{\dagger}T|K_{L,S}\rangle
\right ] \label{eqn:unit1}~~~, \ee
where Im denotes the imaginary part. If we only consider the
two-photon intermediate state, then
\bea \langle l^-l^+|T^{\dagger}T|K_{L,S}\rangle & = &
\sum_{\epsilon, \epsilon'} \int \frac{d^3k}{(2\pi)^3}
\frac{d^3k'}{(2\pi)^3} \frac{1}{2E}\frac{1}{2E'}\langle l^- l^+ |
T^{\dagger} |\gamma(k,\epsilon) \gamma'(k',\epsilon')\rangle
\nonumber \\
&& \hspace{15 ex} \times \langle \gamma(k,\epsilon)
\gamma'(k',\epsilon')|T|K_{L,S}\rangle~~~, \label{eqn:pert} \eea
where $|\gamma(k,\epsilon)\gamma'(k',\epsilon')\rangle$ is a real
two-photon state with $k$ and $k'$, $\epsilon$ and $\epsilon'$
specifying the 4-momenta and 4-polarizations, respectively. Since
the $T$-matrix elements can be expressed as the invariant ${\cal
M}$ matrix elements multiplied by 4-momentum-conserving
$\delta$-functions, Eqs.\ (\ref{eqn:unit1}) and (\ref{eqn:pert})
combine to give
\bea {\rm Im}\left [{\cal M}(K_{L,S} \to l^-l^+) \right ] & = &
\frac{1}{2}\sum_{\epsilon, \epsilon'}
\int\frac{d^3k}{(2\pi)^3}\frac{d^3k'}{(2\pi)^3}\frac{1}{2E}\frac{1}{2E'}
\langle \gamma(k,\epsilon)\gamma'(k',\epsilon')|{\cal
M}|K_{L,S}\rangle \nonumber \\
&& \hspace{2 ex} \langle \gamma(k,\epsilon)\gamma'(k',\epsilon')|
{\cal M} | l^-l^+ \rangle^* (2\pi)^4 \delta^{(4)}(p-k_1-k_2) ,
\label{eqn:unit2} \eea
times an overall $\delta^{(4)}(p-p_1-p_2)$, with $p$, $p_1$ and
$p_2$ being the 4-momenta of $K_{L,S}$, $l^-$ and $l^+$,
respectively. It is expected that the real part of the amplitude
is roughly of the same order as the imaginary part, so that the
actual decay width will exceed the lower bound based on the
imaginary part by only a small factor.

Quite similarly, the imaginary part of the invariant matrix
element ${\cal M}(Z \to {\tilde g}{\tilde g})$ at the one-loop
level can be written as
\be \label{eqn: unit} {\rm Im} \left[{\cal M}(Z \to {\tilde
g}{\tilde g}) \right] = \frac{1}{2} \sum_f \int d\Pi_f {\cal M}(Z
\to f){\cal M}^*({\tilde g}{\tilde g} \to
f)(2\pi)^4\delta^{(4)}(p-\sum_{i=1}^{n_f} p_i) , \ee
where the sum runs over all possible intermediate on-shell states
$f$ and $d\Pi_f = \prod_{i=1}^{n_f}\frac{d^3
p_i}{(2\pi)^3}\frac{1}{2E_i}$ with $n_f$ being the numbers of
particles in state $f$ and $p_i$ being the 3-momenta of the
particles. Since $\sb$ is the lightest supersymmetric particle in
the scenario and all other supersymmetric particles (except $\sg$)
are expected to be heavier than $M_Z/2$, we only need to consider
the cases where $f$ is $b {\bar b}$ and $\sb \bar {\sb}$. The
integral over the phase space $\Pi_f$ can be simplified to the
integral over the solid angle $\Omega$. In the case where the
intermediate state is two particles with equal masses, we have
$\int d\Pi_f (2\pi)^4\delta^{(4)}(p-\sum_{i=1}^{n_f} p_i) =
\frac{v}{32\pi^2}\int d\Omega$, where $v$ is the velocity of the
on-shell intermediate particles.

\section{Amplitudes of the cut diagrams
\label{sec:amps}}

We adopt the spinor convention of Peskin and Schroeder
\cite{peskin}, in which the metric tensor $g_{\mu\nu}={\rm
Diagonal}(1,-1,-1,-1)$ and
\be \gamma^0=\left(\begin{array}{cc} 0 & 1 \\ 1 & 0
\end{array} \right), \hspace{4ex} \gamma^5=\left(\begin{array}{cc} -1 & 0 \\ 0 &
1 \end{array} \right) \hspace{2ex} {\rm and} \hspace{2ex}
\gamma^i=\left(\begin{array}{cc} 0 & \sigma^i \\ -\sigma^i & 0
\end{array} \right),~~ i=1, 2, 3, \nonumber \ee
where $\sigma^i$ are the Pauli matrices. The uncrossed cut Feynman
diagrams that contribute to the imaginary part of the full
amplitude are shown in Fig.~\ref{fig:cut}. The crossed diagrams
with $\sg(k_1) \leftrightarrow \sg(k_2)$ are not shown but should
also be included in the calculation. In the center-of-mass frame
of the $Z$ boson, the 4-momenta of the final gluinos are $k_1 =
(E, \vk)$ and $k_2 = (E, -\vk)$, where $E=M_Z/2$ and ${\vk} = (0,
0, |{\vk}|)$. Suppose $\vk$ is along the $z$-axis and the
polarizations of the $Z$ are quantized along this axis, i.e.,
$\epsilon^\mu = (0, 1, \pm i, 0)/\sqrt{2}$ or $(0, 0, 0, 1)$,
corresponding to helicities $\lambda = \pm 1$ or $0$ respectively.
The 4-momenta of the intermediate bottom quarks are $p_1 = (E,
\vp)$ and $p_2 = (E, -\vp)$, with $\vp = |\vp|(\sin\theta\cos\phi,
\sin\theta\sin\phi, \cos\theta)$. The 4-momenta of the
intermediate bottom squarks are ${\tilde p}_1 = (E, {\tilde
{\vp}})$ and ${\tilde p}_2 = (E, -{\tilde {\vp}})$, with ${\tilde
{\vp}} = |{\tilde {\vp}}|(\sin\theta\cos\phi, \sin\theta\sin\phi,
\cos\theta)$. The Feynman rules for the Majorana fields are given
in a representation independent way in \cite{Haber:1984rc}.

The ${\cal M}$ matrix element for $Z \to b {\bar b}$ is
\be \label{eqn:zbb} {\cal M}(Z \to b {\bar b}) = -\frac{g_W}{2
\cos\theta_W} {\bar u}(p_1) {\not \! \epsilon}(p)(g_L^b
P_L+g_R^bP_R)v(p_2) \delta^{ij}~~~, \ee
where ${\not \! \epsilon} \equiv \epsilon\cdot\gamma$,
$g_L^b=g_V^b+g_A^b=\frac{2}{3}\sin^2\theta_W-1$,
$g_R^b=g_V^b-g_A^b=\frac{2}{3}\sin^2\theta_W$,
$P_L=\frac{1-\gamma^5}{2}$, $P_R=\frac{1+\gamma^5}{2}$,
$\delta^{ij}$ is a Kronecker delta in the quark color indices and
$p = p_1 + p_2$ is the 4-momentum of the $Z$. The Dirac spinors
$u(p_1)$ and $v(p_2)$ can be written as
\bea u^{\ua}(p_1) = \left( \begin{array}{c}
\sqrt{E-|\vp|}\xi^{\ua} \\ \sqrt{E+|\vp|}\xi^{\ua}
\end{array} \right) & \hspace{5ex} & u^{\da}(p_1) = \left(
\begin{array}{c} \sqrt{E+|\vp|}\xi^{\da} \\ \sqrt{E-|\vp|}\xi^{\da}
\end{array} \right) \nonumber \\
v^{\ua}(p_2) = \left( \begin{array}{c} \sqrt{E-|\vp|}\eta^{\ua}
\\ -\sqrt{E+|\vp|}\eta^{\ua}
\end{array} \right) & \hspace{5ex} & v^{\da}(p_2) = \left(
\begin{array}{c} \sqrt{E+|\vp|}\eta^{\da}
\\ -\sqrt{E-|\vp|}\eta^{\da}
\end{array} \right)~~,
\eea
where the arrows $\ua$ and $\da$ denote spin up and spin down
along $\vp$, respectively;
\bea \xi^\uparrow = \left( \begin{array}{c} \cos \frac{\theta}{2}
\\ e^{i\phi}\sin \frac{\theta}{2} \end{array} \right)
& \hspace{7ex} & \xi^\downarrow = \left(
\begin{array}{c} -e^{-i\phi}\sin \frac{\theta}{2} \\ \cos
\frac{\theta}{2} \end{array} \right) \nonumber \\
\eta^\uparrow = \left(\begin{array}{c}
-\sin \frac{\theta}{2} \\
e^{i\phi}\cos \frac{\theta}{2} \end{array} \right) & \hspace{7ex}
& \eta^\downarrow = \left(\begin{array}{c} e^{-i\phi}\cos
\frac{\theta}{2} \\ \sin \frac{\theta}{2}
\end{array} \right)~~~.  \nonumber \eea
We then have
\bea {\cal M}(Z \to b^\ua {\bar b}^\ua) & = & (0,
i\sin\phi+\cos\phi\cos\theta,
-i\cos\phi+\sin\phi\cos\theta, -\sin\theta)\cdot \epsilon(p) \nonumber \\
& & \hspace{15 ex}  \times \frac{g_W}{2 \cos\theta_W}
\left[(E-|{\vp}|)g_L^b + (E+|{\vp}|)g_R^b \right] \delta^{ij} \nonumber \\
{\cal M}(Z \to b^\da {\bar b}^\da) & = & (0,
-i\sin\phi+\cos\phi\cos\theta,
i\cos\phi+\sin\phi\cos\theta, -\sin\theta)\cdot \epsilon(p) \nonumber \\
& & \hspace{15 ex}  \times \frac{g_W}{2 \cos\theta_W}
\left[(E+|{\vp}|)g_L^b + (E-|{\vp}|)g_R^b \right] \delta^{ij} \nonumber \\
{\cal M}(Z \to b^\ua {\bar b}^\da) & = & \frac{g_W m_b}{2
\cos\theta_W}e^{-i\phi}\left[ g_L^b
(-1,\sin\theta\cos\phi,\sin\theta\sin\phi,\cos\theta)
\right. \nonumber \\
& & \hspace{15 ex} \left.
+g_R^b(1,\sin\theta\cos\phi,\sin\theta\sin\phi,\cos\theta)\right]\cdot
\epsilon(p) \delta^{ij} \nonumber \\
{\cal M}(Z \to b^\da {\bar b}^\ua) & = & -\frac{g_W m_b}{2
\cos\theta_W}e^{i\phi}\left[ g_L^b
(1,\sin\theta\cos\phi,\sin\theta\sin\phi,\cos\theta)
\right. \nonumber \\
& & \hspace{15 ex} \left.
+g_R^b(-1,\sin\theta\cos\phi,\sin\theta\sin\phi,\cos\theta)\right]\cdot
\epsilon(p)\delta^{ij} \nonumber  \eea
For $\sin^2\theta_W=0.2311$ and without top quark corrections, the
partial decay width for $Z$ to decay into massless $b{\bar b}$ is
then $\frac{G_F M_Z^3}{4\sqrt{2}\pi}[(g_L^b)^2+(g_R^b)^2] = 368$
MeV.

Now we consider $b {\bar b} \to \sg \sg$ via exchange of a $\sb$
or $\sb'$, the matrix element for which is denoted ${\cal M}(b
{\bar b} \to \sg \sg)$ or ${\cal M}'(b {\bar b} \to \sg \sg)$,
respectively. We have ${\cal M}(b {\bar b} \to \sg \sg)={\cal
M}^{(1)}(b {\bar b} \to \sg \sg)+{\cal M}^{(2)}(b {\bar b} \to \sg
\sg)$, with
\bea {\cal M}^{(1)}(b {\bar b} \to \sg \sg) & = & -2
g_s^2\frac{(t^b t^a)_{ji}}{(p_1-k_1)^2-m_{\sb}^2}{\bar {\tilde
u}}(k_1) \left(P_L\sin\theta_{\sb}-P_R\cos\theta_{\sb}
\right) u(p_1) \nonumber \\
& &  \hspace{20 ex}  {\bar v}(p_2) \left(P_R\sin\theta_{\sb} -
P_L\cos\theta_{\sb} \right) {\tilde v}(k_2) , \label{eqn:uncr}\\
{\cal M}^{(2)}(b {\bar b} \to \sg \sg) & = & -2 g_s^2 \frac{(t^a
t^b)_{ji}}{(p_1-k_2)^2-m_{\sb}^2} {\tilde v}^T(k_2)C^{-1}
\left(P_L\sin\theta_{\sb}-P_R\cos\theta_{\sb}
\right) u(p_1) \nonumber \\
& &  \hspace{20 ex}  {\bar v}(p_2) \left(P_R\sin\theta_{\sb} -
P_L\cos\theta_{\sb} \right) C {\bar {\tilde u}}^T(k_1) ,
\label{eqn:cr}\eea
where the superscript (1) denotes the uncrossed diagram and (2)
the crossed diagram;  $a, b$ and $i, j$ are the color indices of
the gluinos and the quarks, respectively; $t^a$ are the
fundamental representation matrices of $SU(3)$;
$C=i\gamma^0\gamma^2$ is the charge conjugate matrix. It can be
easily verified that $u(p, s)=C{\bar v}^T(p, s)$ and $v(p,
s)=C{\bar u}^T(p, s)$, where $T$ means ``transpose''. The Majorana
spinors ${\tilde u}(k_1)$ and ${\tilde v}(k_2)$ also satisfy these
relations \cite{Haber:1984rc}. Thus we can immediately write
\bea {\tilde u}^{\ua}(k_1) = \left( \begin{array}{c}
\sqrt{E-|\vk|}\zeta_+ \\ \sqrt{E+|\vk|}\zeta_+
\end{array} \right) & \hspace{5ex} & {\tilde u}^{\da}(k_1) =
\left( \begin{array}{c} \sqrt{E+|\vk|}\zeta_- \\
\sqrt{E-|\vk|}\zeta_-
\end{array} \right) \nonumber \\
{\tilde u}^{\ua}(k_2) = \left( \begin{array}{c}
\sqrt{E+|\vk|}\zeta_+ \\ \sqrt{E-|\vk|}\zeta_+
\end{array} \right) & \hspace{5ex} & {\tilde u}^{\da}(k_2) =
\left( \begin{array}{c} -\sqrt{E-|\vk|}\zeta_- \\
-\sqrt{E+|\vk|}\zeta_-
\end{array} \right) \nonumber \\
{\tilde v}^{\ua}(k_1) = \left( \begin{array}{c}
\sqrt{E+|\vk|}\zeta_- \\ -\sqrt{E-|\vk|}\zeta_-
\end{array} \right) & \hspace{5ex} & {\tilde v}^{\da}(k_1) = \left(
\begin{array}{c}
-\sqrt{E-|\vk|}\zeta_+ \\ \sqrt{E+|\vk|}\zeta_+
\end{array} \right) \nonumber \\
{\tilde v}^{\ua}(k_2) = \left( \begin{array}{c}
\sqrt{E-|\vk|}\zeta_- \\ -\sqrt{E+|\vk|}\zeta_-
\end{array} \right) & \hspace{5ex} & {\tilde v}^{\da}(k_2) = \left(
\begin{array}{c}
\sqrt{E+|\vk|}\zeta_+
\\ -\sqrt{E-|\vk|}\zeta_+
\end{array} \right)~~,
\eea
with $\zeta_+=\left(\begin{array}{c} 1 \\ 0
\end{array} \right)$ and $\zeta_-=\left(\begin{array}{c} 0 \\ 1
\end{array} \right)$. Here the arrows $\ua$ and $\da$ denote spin up
and spin down along $\vk$ (i.e., the $z$-axis), respectively.
Since ${\tilde v}^T(k_2)C^{-1}=-{\bar {\tilde u}}(k_2)$ and $C
{\bar {\tilde u}}^T(k_1)={\tilde v}(k_1)$, Eq.\ (\ref{eqn:cr}) can
alternatively be obtained from Eq.\ (\ref{eqn:uncr}) by
interchanging $k_1$ and $k_2$ and adding an overall minus sign.
The helicities of the final gluinos are determined by $\lambda$,
the initial helicity of the $Z$. For $\lambda = 1$, both gluinos
have spin up in the $z$-direction, while for $\lambda = -1$, both
have spin down in the $z$-direction. For $\lambda = 0$, one of
them has spin up and the other has spin down in the $z$-direction.
One expects $|{\rm Im}{\cal M}(Z^{\da} \to \sg^{\da}
\sg^{\da})|=|{\rm Im}{\cal M}(Z^{\ua} \to \sg^{\ua} \sg^{\ua})|$,
because the two processes are related by mirror symmetry. One also
expects ${\rm Im}{\cal M}(Z^{(0)} \to \sg^{\ua} \sg^{\da})=0$,
because these final gluinos have the same helicities and should
therefore be excluded by the Pauli principle. The matrix element
${\cal M}'(b {\bar b} \to \sg \sg)$ can be obtained from ${\cal
M}(b {\bar b} \to \sg \sg)$ by replacing $m_{\sb}$,
$\sin\theta_{\sb}$ and $\cos\theta_{\sb}$ with $m_{\sb'}$,
$\cos\theta_{\sb}$ and $-\sin\theta_{\sb}$, respectively.

Now consider the diagram in Fig.~\ref{fig:cut} (b) and a similar
diagram with $\sg(k_1) \leftrightarrow \sg(k_2)$, where the
intermediate state is a pair of scalar quarks ($\sb$ and $\bar
{\sb}$). The tree-level $Z\sb\bar {\sb}$ coupling is proportional
to $g_L^b\sin^2\theta_{\sb}+ g_R^b\cos^2\theta_{\sb}$, so a mixing
angle of $\theta_{\sb} = \arcsin{\sqrt{2\sin^2\theta_W/3}} \simeq
23^\circ$ or $157^\circ$ will make it vanish. A weak $Z\sb\bar
{\sb}$ coupling is assumed \cite{Berger:2000mp} to satisfy the
tight constraints imposed by precision measurements at the $Z$
peak. Consequently the contribution of the $\sb\bar {\sb}$
intermediate state to ${\rm Im}{\cal M}(Z \to \sg \sg)$ should
also be small. However, to see how the two types of diagrams shown
in Fig.~\ref{fig:cut} interfere with each other, we take
$\theta_{\sb}$ to be a free parameter. For the first part of the
cut diagram [Fig.~\ref{fig:cut} (b)], we have
\be \label{eqn:zsbsb} {\cal M}(Z \to \sb \bar {\sb}) =
-\frac{g_W}{2 \cos\theta_W}\left[g_L^b\sin^2\theta_{\sb}+ g_R^b
\cos^2\theta_{\sb}\right] ({\tilde p}_1-{\tilde
p}_2)^\mu\epsilon_\mu(p) \delta^{ij} ,\ee
where $i$ and $j$ are the squark color indices. The sign
discrepancy mentioned in the Introduction can be traced to the
relative sign between Eqs.\ (\ref{eqn:zbb}) and  (\ref{eqn:zsbsb})
\cite{Berge:2002ev}. The current sign in Eq.\ (\ref{eqn:zsbsb}) is
consistent with the Feynman rules in Ref.~\cite{Haber:1984rc}. We
will argue in favor of this sign from another point of view in
Section \ref{sec:fdw}. For the other part of the diagram,
\bea {\cal M}^{(1)}(\sb \bar {\sb} \to \sg \sg) & = & 2g_s^2\frac{
(t^b t^a)_{ji}}{({\tilde p}_1-k_1)^2-m_b^2}{\tilde v}^T(k_2)
C^{-1} \left[P_L\sin\theta_{\sb} - P_R\cos\theta_{\sb}
\right]  \nonumber \\
&& ({\not \! {\tilde p}}_1-{\not \! k}_1+m_b) \left[P_R\sin\theta_{\sb} -
P_L\cos\theta_{\sb} \right] C{\bar {\tilde u}}^T(k_1) \\
{\cal M}^{(2)}(\sb \bar {\sb} \to \sg \sg) & = & 2g_s^2\frac{(t^a
t^b)_{ji}}{({\tilde p}_1-k_2)^2-m_b^2} {\bar {\tilde
u}}(k_1)\left[P_L\sin\theta_{\sb} - P_R\cos\theta_{\sb} \right]
\nonumber \\
&& ({\not \! {\tilde p}}_1-{\not \! k}_2+m_b)
\left[P_R\sin\theta_{\sb} - P_L\cos\theta_{\sb} \right] {\tilde
v}(k_2) ,\eea
where (1) denotes the uncrossed diagram and (2) the crossed
diagram. The relevant matrix elements for $\lambda=1$ are
presented in the Appendix.

\section{Lower bound on $\Gamma(Z \to \sg \sg)$
\label{sec:lb}}

\begin{figure}
\centerline{\epsfysize = 6 in
\epsffile{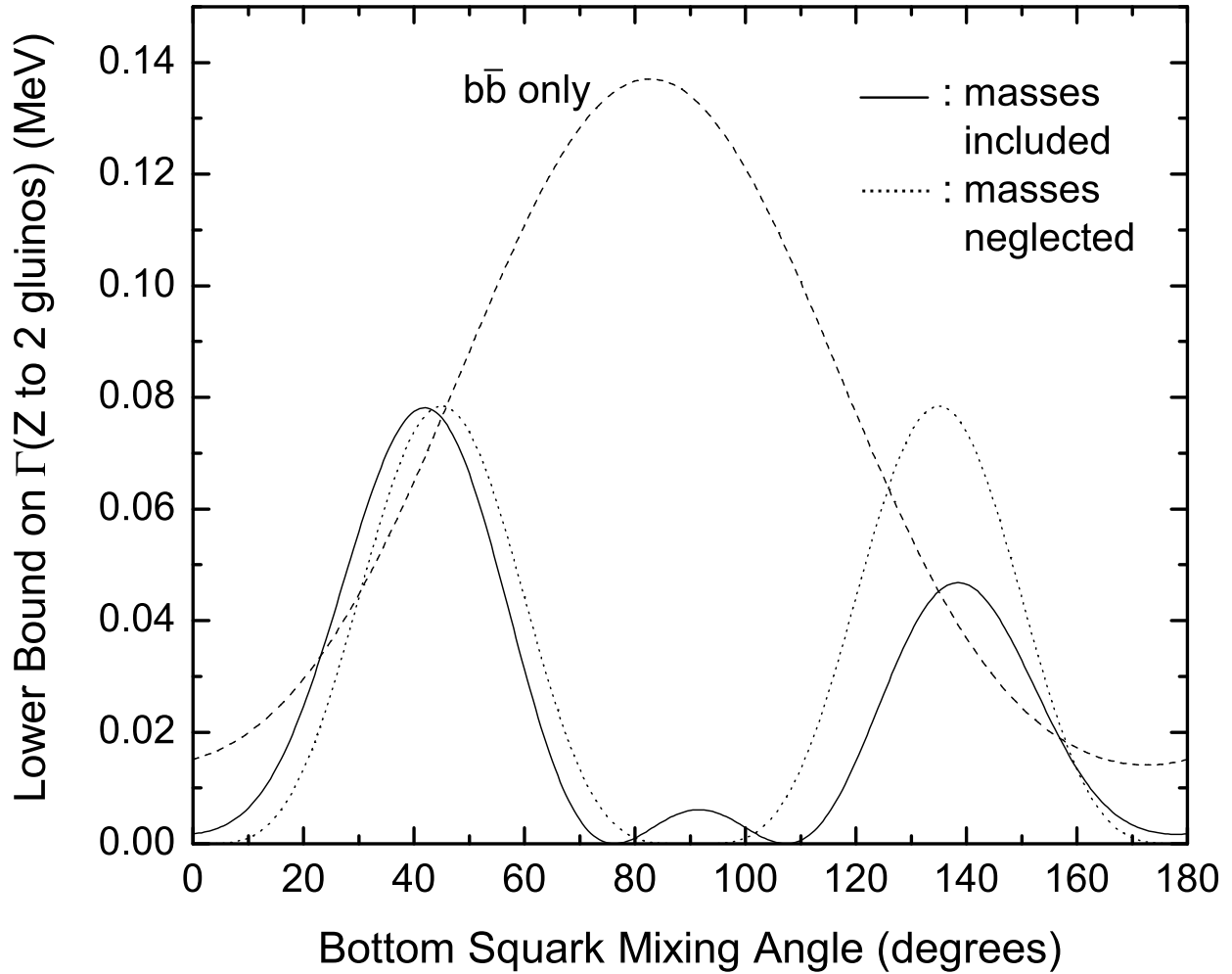}
}
\caption[]{\label{fig:lb}
   Lower bound on $\Gamma(Z \to \sg \sg)$ as a function of the
bottom squark mixing angle $\theta_{\sb}$. Solid curve: $m_b=4.1$
GeV, $m_{\sb}=4.5$ GeV, $m_{\sg}=15$ GeV, $m_{\sb'}=170$ GeV;
dashed curve: only the contribution from the $b\bar b$
intermediate state is included with the same set of masses; dotted
curve: $m_b= m_{\sb}= m_{\sg}=0$ and $m_{\sb'}=\infty$. }
\end{figure}

Now we are ready to put things together and obtain a lower bound
on $\Gamma(Z \to \sg \sg)$. First we consider an extreme case with
$m_b=m_{\sb}=m_{\sg}=0$ and $m_{\sb'}=\infty$. In this limit, the
product ${\cal M}(Z \to f){\cal M}(f \to \sg \sg)$ has an angular
dependence of either $(1+\cos\theta)$ or $(1-\cos\theta)$.
However, the $\cos\theta$ term does not contribute to the
imaginary part of the full amplitude, because integrating it over
the solid angle $\Omega$ gives zero. Note that ${\rm tr}(t^a
t^b)={\rm tr}(t^b t^a)=\delta^{ab}/2$. The only nonvanishing
amplitudes are
\begin{eqnarray*}
{\cal M}(Z^\ua \to b{\bar b})*{\cal M}(b{\bar b} \to \sg^\ua
\sg^\ua) & = & \delta^{ab} r_W (g_A^b - g_V^b\cos2\theta_{\sb}) \\
{\cal M}(Z^\da \to b{\bar b})*{\cal M}(b{\bar b} \to \sg^\da
\sg^\da) & = & -\delta^{ab} r_W (g_A^b - g_V^b\cos2\theta_{\sb}) \\
{\cal M}(Z^\ua \to \sb\bar {\sb})*{\cal M}(\sb\bar {\sb} \to
\sg^\ua \sg^\ua) & = & \delta^{ab} r_W (g_V^b - g_A^b
\cos2\theta_{\sb}) \cos2\theta_{\sb} \\
{\cal M}(Z^\da \to \sb\bar {\sb})*{\cal M}(\sb\bar {\sb} \to
\sg^\da \sg^\da) & = & -\delta^{ab} r_W (g_V^b - g_A^b
\cos2\theta_{\sb}) \cos2\theta_{\sb} ~~~,
\end{eqnarray*}
where the asterisks imply that we have integrated over the phase
spaces and summed over all intermediate helicity states,
$r_W=\frac{M_Z g_W g_s^2}{2 \sqrt{2}\cos\theta_W}$. From the above
equations we can see that the imaginary parts of diagrams (a) and
(b) in Fig.~\ref{fig:cut} interfere destructively if $\sb$ is more
left-handed ($45^\circ<\theta_{\sb}<135^\circ$) or dominantly
right-handed ($\theta_{\sb}<23^\circ$ or
$\theta_{\sb}>157^\circ$); the contribution of diagram (b) remains
negligible in the neighborhood of the decoupling angle ($23^\circ$
or $157^{\circ}$). The imaginary parts of the amplitudes are
\bea {\rm Im}{\cal M}(Z^{\ua} \to \sg^{\ua} \sg^{\ua}) & = & -
{\rm Im}{\cal M}(Z^{\da} \to \sg^{\da} \sg^{\da})
\label{eqn:mirror}\\
& = & \delta^{ab} r_W (g_L^b-g_R^b) \sin^2\theta_{\sb}
\cos^2\theta_{\sb}/(8\pi) ~~~. \nonumber \eea
The relation (\ref{eqn:mirror}) also holds when all the particles
have a finite mass. The lower bound on $\Gamma(Z \to \sg \sg)$ in
the limit $m_b=m_{\sb}=m_{\sg}=0$ and $m_{\sb'}=\infty$ can be
expressed as a ratio
\bea \frac{\Gamma(Z \to \sg \sg)}{\Gamma(Z \to b{\bar b})} & \ge &
\frac{1}{2}\frac{[{\rm Im}{\cal M}(Z^{\ua} \to \sg^{\ua}
\sg^{\ua})]^2+[{\rm Im}{\cal M}(Z^{\da} \to \sg^{\da}
\sg^{\da})]^2}{[{\rm Im}{\cal M}(Z^{\ua} \to b^{\ua} {\bar
b}^{\ua})]^2+[{\rm Im}{\cal M}(Z^{\da} \to
b^{\da} {\bar b}^{\da})]^2} \nonumber \\
& = & \frac{\alpha_s^2}{6}
\frac{(g_L^b-g_R^b)^2\sin^4\theta_{\sb}\cos^4\theta_{\sb}}{(g_L^b)^2
+(g_R^b)^2}~~~. \label{eqn:ratio} \eea
The factor of $1/2$ comes in because the final gluinos are
identical. Taking $\Gamma(Z \to b{\bar b})=368$ MeV, we plot the
lower bound on the decay width $\Gamma(Z \to \sg \sg)$ as a
function of the bottom squark mixing angle $\theta_{\sb}$ in
Fig.~\ref{fig:lb} (dotted curve). When all the masses are finite,
we can no longer ignore the $\cos\theta$ terms, because the
denominators of the propagators are no longer of the form $\sim(1
\pm \cos\theta)$, which previously cancelled with the same factors
in the numerators of the amplitudes and gave only linear terms in
$\cos\theta$. However, it is still not hard to perform the
integration over the angles. Define
\begin{eqnarray}
I_\pm(x,y,z) & = & \frac{1}{2}\int_0^\pi
\frac{(1\pm\cos\theta)^2}{x^2 + y^2 + z^2 + 2 x y\cos\theta}
\sin\theta d\theta \nonumber \\
I_0(x,y,z) & = & \int_0^\pi \frac{\sin^2\theta}{x^2 + y^2 + z^2 +
2 x y\cos\theta} \sin\theta d\theta~~~ \label{eqn:int},
\end{eqnarray}
and let $c_\pm = I_\pm(v_b, v_{\sg}, r_{\sb})$, $c'_\pm =
I_\pm(v_b, v_{\sg}, r_{\sb'})$, $c_0 =I_0(v_b, v_{\sg}, r_{\sb})$,
$c'_0 =I_0(v_b, v_{\sg}, r_{\sb'})$, ${\tilde c}_0=I_0(v_{\sb},
v_{\sg}, r_b)$, where $r_i = 2m_i/M_Z$ ($i=b,\sb,\sb',\sg$), $v_i
= \sqrt{1-r_i^2}$ is the ``velocity'' of an on-shell particle $i$
($i=b,\sb,\sg$). The lower bound can then be written as
\be \label{eqn:lb} \Gamma(Z \to \sg \sg) \ge \frac{G_F M_Z^3
\alpha_s^2} {96\sqrt{2}\pi} ({\cal A}_1 v_b + {\cal A}'_1 v_b +
{\cal A}_2 v_{\sb})^2 v_{\sg}~~~, \ee
where, up to a common factor of proportionality, ${\cal A}_1 v_b$
and ${\cal A}'_1 v_b$ are the imaginary parts of the amplitudes
for $Z \to (b\bar b)^* \to \sg\sg$ via exchange of a $\sb$ and
$\sb'$, respectively; ${\cal A}_2 v_{\sb}$ is the imaginary part
of the amplitude for $Z \to (\sb\bar {\sb})^* \to \sg\sg$. We have
\begin{eqnarray*} {\cal A}_1 & = & c_1 (g_A^b-g_V^b\cos2\theta_{\sb})
- g_A^b(c_2  + c_3 \sin2\theta_{\sb}) \\
{\cal A}'_1 & = & c'_1 (g_A^b+g_V^b\cos2\theta_{\sb}) - g_A^b(c'_2
- c'_3 \sin2\theta_{\sb}) \\
{\cal A}_2 & = & \tilde c_0 v_{\sb}^2 v_{\sg}
(g_V^b-g_A^b\cos2\theta_{\sb}) \cos2\theta_{\sb}~~~,
\end{eqnarray*}
where $c_1=c_-(v_{\sg}-v_b) + c_+(v_{\sg}+v_b) + c_0 r_b^2
v_{\sg}$,~~$c_2 =(c_- + c_+ + c_0)r_b^2 v_{\sg}$,~~$c_3 = (c_+ -
c_-)v_b r_b r_{\sg}$; $c'_1$, $c'_2$ and $c'_3$ are defined
similarly, with $c_\pm$ and $c_0$ all primed. All these quantities
only depend on the masses. As $m_b$, $m_{\sb}$ and $m_{\sg}$ go to
zero and $m_{\sb'}$ goes to infinity, $c_1 \to 1$, $\tilde c_0
v_{\sb}^2 v_{\sg} \to 1$, $c_2 \to 0$, $c_3 \to 0$, $c'_1 \to 0$,
$c'_2 \to 0$ and $c'_3 \to 0$. Eq.\ (\ref{eqn:ratio}) is thus
recovered. As far as the imaginary part of the amplitude is
concerned, Eq.\ (\ref{eqn:lb}) agrees with
Ref.~\cite{Djouadi:1994js} in the limit $m_{\sg}=0$ except for the
mentioned sign discrepancy. Compared to
Ref.~\cite{Djouadi:1994js}, this unitarity calculation not only
takes into account a nonzero gluino mass but also is much simpler.
Firstly, there are no singularities to remove. Secondly, one only
has to evaluate a few elementary integrals in Eq.\ (\ref{eqn:int})
instead of the much more difficult one-loop two- and three-point
functions.

The lower bound is plotted in Fig.~\ref{fig:lb} (solid curve) as a
function of $\theta_{\sb}$ for a specific set of values for the
masses: $m_b=4.1$ GeV, $m_{\sb}=4.5$ GeV, $m_{\sb'}=170$ GeV,
$m_{\sg}=15$ GeV. The small peak around $90^\circ$ disappears if
the gluino has a small mass (e.g., $m_{\sg} \le 7$ GeV) such that
the sign of the sum ${\cal A}_1 v_b + {\cal A}'_1 v_b + {\cal A}_2
v_{\sb}$ does not change over $0^\circ \le \theta_{\sb} \le
180^\circ$. If $\sg$ were massless, the lower bound as a function
of $0^\circ \le \theta_{\sb}\le 180^{\circ}$ would have two
identical peaks (see the dashed curve in Fig.~\ref{fig:fw}). Due
to destructive interference between diagrams (a) and (b) for
$45^\circ<\theta_{\sb}<135^\circ$, $\theta_{\sb}<23^\circ$ and
$\theta_{\sb}>157^\circ$, the lower bound is smaller than the
contribution from the $b\bar b$ intermediate state alone
(Fig.~\ref{fig:lb}, dashed curve) in these ranges.

\section{Implications for the full decay width $\Gamma(Z \to \sg \sg)$
\label{sec:fdw}}

Let us first examine under what conditions the decay amplitude
vanishes. We will see that Kane and Rolnick's claim \cite{Kane:xp}
contradicts our unitarity calculation. Under their assumption,
$m_{\tilde q} = m_{\tilde q'} = m_q$ and $\theta_{\tilde q} = 0$,
so that $c_i=c'_i$ ($i=1, 2, 3$) and ${\tilde c}_0 = {\tilde c'}_0
\equiv I_0(v_{\sb'}, v_{\sg}, r_b) = \pm c_0$. We regard the
relative sign ($\pm$) between diagrams (a) and (b) as undecided
and only consider the contribution of the third generation, i.e.,
$q=t, b$. Define ${\cal A}'_2 \equiv -\tilde c'_0 v_{\sb'}^2
v_{\sg} (g_V^b+g_A^b\cos2\theta_{\sb}) \cos2\theta_{\sb}$. Then
the imaginary part of the amplitude is proportional to ${\cal A}_1
+ {\cal A}'_1 + {\cal A}_2 + {\cal A}'_2 = 2[(c_+ + c_- \mp c_0)
v_b^2 v_{\sg} + (c_+ - c_-) v_b]g_A^b$, which is nonzero for
either sign. Therefore, the amplitude does $\it not$ vanish under
Kane and Rolnick's conditions. For the imaginary part to be zero,
we must have $m_t = m_b$ and $m_{\st} = m_{\st'} = m_{\sb} =
m_{\sb'}$ and similar mass degeneracies in the other two
generations. (Of course, the imaginary part of the amplitude
automatically vanishes if all quark and squark mass eigenstates
are heavier than $M_Z/2$.) This is true whether the relative sign
between diagrams (a) and (b) is $+$ or $-$ and is consistent with
Refs.~\cite{Berge:2002ev, Kileng:1994vc, Djouadi:1994js}.

Let us now investigate whether the imaginary part of the amplitude
computed in the previous section can give us some hint how loop
divergences cancel and reasonably small decay widths for $Z \to
\sg \sg$ can be obtained. Without actually calculating the full
amplitude, we should be able to recover part of it from the
imaginary part. For simplicity we take $m_{\sg}=0$ and consider
only diagram (b) in Fig.~\ref{fig:cut}. The imaginary part of the
amplitude for this diagram is proportional to ${\cal A}_2 v_{\sb}
= \tilde c_0 v_{\sb}^3 (g_V^b-g_A^b\cos2\theta_{\sb})
\cos2\theta_{\sb}$. We have
$$ \frac{\pi}{2} \tilde c_0 v_{\sb}^3 =\left[\frac{1}{2} -
\frac{m_{\sb}^2-m_b^2}{s}\right] v_{\sb} \pi + \left[m_b^2 +
\frac{(m_{\sb}^2-m_b^2)^2}{s}\right] \frac{\pi}{s}
\log\frac{(m_{\sb}^2 - m_b^2)/s -(1-v_{\sb})/2}{(m_{\sb}^2 - m_b^2)/s
-(1+v_{\sb})/2} ,$$
where $s=M_Z^2$. For $m_{\sb}< M_Z/2$, one can check that
\begin{eqnarray*} {\rm Im} B_0(s, m_{\sb}, m_{\sb}) & = & v_{\sb} \pi  \\
{\rm Im} C_0(s, m_{\sb}, m_{\sb}, m_b) & = & \frac{\pi}{s}
\log\frac{(m_{\sb}^2 - m_b^2)/s -(1-v_{\sb})/2}{(m_{\sb}^2 - m_b^2)/s
-(1+v_{\sb})/2}~~~,
\end{eqnarray*}
where $B_0$ and $C_0$ are the scalar one-loop two- and three-point
functions \cite{Djouadi:1994js, private, 'tHooft:1978xw,
Drees:1991zk}, respectively. If we define
\begin{eqnarray*} A(m_1, m_2, m_3) & = & \left[\frac{1}{2} -
\frac{(m_1^2-m_3^2)+(m_2^2-m_3^2)}{2s}\right] B_0(s, m_1, m_2) \\
& & + \left[m_3^2 + \frac{(m_1^2 - m_3^2)(m_2^2-m_3^2)}{s}\right]
C_0(s, m_1, m_2, m_3)~~~,
\end{eqnarray*}
then by analyticity and symmetry, the following terms should be
part of the full amplitude contributed by type (b) diagrams in
Fig.~\ref{fig:cut},
\begin{eqnarray}
&& A(m_{\sb},m_{\sb},m_b) (g_V^b-g_A^b\cos2\theta_{\sb})
\cos2\theta_{\sb} - A(m_{\sb'},m_{\sb'},m_b)
(g_V^b+g_A^b\cos2\theta_{\sb}) \cos2\theta_{\sb} \nonumber
\\
&&-A(m_{\sb},m_{\sb'},m_b)2g_A^b\sin^22\theta_{\sb} +
A(m_{\st},m_{\st},m_t) (g_V^t-g_A^t\cos2\theta_{\st})
\cos2\theta_{\st} \nonumber \\
&& -A(m_{\st'},m_{\st'},m_t) (g_V^t+g_A^t\cos2\theta_{\st})
\cos2\theta_{\st}-A(m_{\st},m_{\st'},m_t)2g_A^t\sin^22\theta_{\st}
\label{eqn:sum}~~~.
\end{eqnarray}
Here the $\sin^22\theta_{\tilde q}$ terms can be obtained by
repeating the unitarity calculation with $f=\sb {\bar {\sb'}}$.
Alternatively, they can easily be guessed if we note that the
above terms should sum up to zero for $m_t=m_b$ and $m_{\st} =
m_{\st'} = m_{\sb} = m_{\sb'}$. Those terms in Eq.\
(\ref{eqn:sum}) are expected to be the only ones relevant to our
argument \cite{Djouadi:1994js}.

The top and bottom squark masses and mixing angles are determined
by the following mass matrices \cite{Djouadi:1994js}
\bea M_{\st}^2 & = & \left( \begin{array}{cc} m_{\st_L}^2 + m_t^2
+ g_L^t s \cos 2\beta/2 & -m_t (A_t + \mu \cot\beta) \\ -m_t (A_t
+ \mu \cot\beta) & m_{\st_R}^2 + m_t^2 - g_R^t s \cos
2\beta/2 \end{array}\right) \\
M_{\sb}^2 & = & \left( \begin{array}{cc} m_{\sb_L}^2 + m_b^2 +
g_L^b s \cos 2\beta/2 & -m_b (A_b + \mu \tan\beta) \\ -m_b (A_b +
\mu \tan\beta) & m_{\sb_R}^2 + m_b^2 - g_R^b s \cos 2\beta/2
\end{array}\right)~~~,
\eea
where $m_{\st_L}^2, m_{\st_R}^2, m_{\sb_L}^2, m_{\sb_R}^2$ are
soft SUSY breaking masses; $\tan\beta$ is the ratio of the vacuum
expectation values of the two neutral Higgs fields in the MSSM;
$A_t, A_b$ denote the trilinear Higgs-stop, -sbottom couplings,
respectively; and $\mu$ is the Higgsino mass parameter. $SU(2)$
gauge invariance leads to $m_{\st_L}^2 = m_{\sb_L}^2$. It is
reasonable to assume that we should get a sensible decay width
$\Gamma(Z \to \sg\sg)$ for each specific set of parameters. In
particular, we can choose very large values for $m_{\st},
m_{\st'}, m_{\sb}, m_{\sb'}$ and should find a tiny $\Gamma(Z \to
\sg \sg)$ for some mixing angles $\theta_{\st}$ and
$\theta_{\sb}$. For simplicity we assume $A_t + \mu \cot\beta =
A_b + \mu \tan\beta = 0$ so that there is no $L$-$R$ squark
mixing, and
\begin{eqnarray*}
m_{\st_L}^2 + g_L^t s \cos 2\beta/2 & = & m_{\st_R}^2 - g_R^t
s \cos 2\beta/2 \\
m_{\st_L}^2 + g_L^b s \cos 2\beta/2 & = & m_{\sb_R}^2 - g_R^b s
\cos 2\beta/2
\end{eqnarray*}
so that $m_{\st} = m_{\st'}$, $m_{\sb} = m_{\sb'}$. The mass
difference between the top and bottom squarks is then
\begin{displaymath} m_{\st} - m_{\sb} = \sqrt{m_{\st_L}^2 + m_t^2
 + g_L^t s \cos 2\beta/2} - \sqrt{m_{\st_L}^2 + m_b^2 + g_L^b s \cos
2\beta/2} \approx \frac{\delta m^2}{m_{\st_L}}~~~, \end{displaymath}
with $\delta m^2 \equiv [2(m_t^2 - m_b^2) + (g_L^t-g_L^b) s \cos
2\beta]/4$. The sum in Eq.\ (\ref{eqn:sum}) becomes simply
\be A(m_{\sb}, m_{\sb}, m_b) - A(m_{\st}, m_{\st}, m_t) \equiv
\delta A~~~. \ee
In the heavy squark limit,
\begin{eqnarray*} B_0(s, m_{\tilde q}, m_{\tilde q}) & = & \Delta
+ \log \frac{m_{\tilde q}^2}{\nu^2} \\
C_0(s, m_{\tilde q}, m_{\tilde q}, m_q) & = & -\frac{1}{m_{\tilde
q}^2 - m_q^2} + \frac{m_q^2}{(m_{\tilde q}^2 - m_q^2)^2} \log
\frac{m_{\tilde q}^2}{m_q^2}~~~,
\end{eqnarray*}
where $\Delta$ denotes the ultraviolet divergent part of $B_0$ and
$\nu$ is the renormalization scale. We can see that the $B_0$ and
$C_0$ terms in $A(m_{\tilde q}, m_{\tilde q}, m_q)$ vary as
$-(m_{\tilde q}^2-m_q^2)[\log m_{\tilde q}^2/\nu^2]/s$ and
$-(m_{\tilde q}^2-m_q^2)/s$, respectively. So the leading term in
$\delta A$ comes from $B_0$ and varies as
$$ [(m_{\st}^2-m_t^2) \log m_{\st}^2 - (m_{\sb}^2-m_b^2) \log 
m_{\sb}^2]/s \sim [(g_L^t-g_L^b) \cos 2\beta \log m_{\st_L}^2]/2~~~. $$
Thus the sum in Eq.\ (\ref{eqn:sum}) is logarithmically divergent
as $m_{\st_L}$ goes to infinity. Therefore, we cannot get
reasonable decay widths if this divergence is supposed to cancel
within an isodoublet. So we do not agree with Djouadi and Drees'
claim \cite{Djouadi:1994js} that one can get meaningful results by
summing over a complete isodoublet. The only other way out is for
the divergence to cancel for each weak isospin partner. Note that
the ultraviolet divergent parts of the $B_0$ terms in Eq.\
(\ref{eqn:sum}) not only can cancel within an isodoublet, but also
can cancel for each weak isospin partner. Indeed, if the relative
sign between diagrams (a) and (b) in Ref.~\cite{Djouadi:1994js} is
reversed, one finds that both ultraviolet divergences and the
divergences in the heavy squark limit (as discussed above) cancel
separately for the top and bottom sectors. To understand how the
latter divergences cancel, repeat the same reasoning as that
leading to Eq.\ (\ref{eqn:sum}) for type (a) diagrams in
Fig.~\ref{fig:cut} and observe that those diagrams should
contribute the following terms,
\bea & & \pm \left[C(m_b, m_b, m_{\sb})
(g_A^b-g_V^b\cos2\theta_{\sb}) + C(m_b, m_b,
m_{\sb'})(g_A^b+g_V^b\cos2\theta_{\sb})
\right. \nonumber \\
& & \hspace{2ex}\left. C(m_t, m_t, m_{\st})
(g_A^t-g_V^t\cos2\theta_{\st}) + C(m_t, m_t, m_{\st'})
(g_A^t+g_V^t\cos2\theta_{\st}) \right]~~~, \label{eqn:a} \eea
where again we keep an undecided relative sign and
$$ C(m_1, m_1, m_2) \equiv \left[m_1^2 + \frac{(m_2^2 -
m_1^2)^2}{s}\right] C_0(s, m_1, m_1, m_2)~~~. $$
For $m_{\st} = m_{\st'}$ and $m_{\sb} = m_{\sb'}$, Eq.\
(\ref{eqn:a}) becomes
\be \pm [C(m_t, m_t, m_{\st}) - C(m_b, m_b, m_{\sb})] \equiv \pm
\delta C~~~. \ee
One can check that when $m_{\tilde q}$ becomes large,
$$ {\rm Re} C_0(s, m_q, m_q, m_{\tilde q}) \sim -\frac{1}{m_{\tilde
q}^2 - m_q^2} \log \frac{m_{\tilde q}^2}{s}~~~. $$
Thus the leading term in ${\rm Re}\delta C$ is exactly the same as
the leading term in $\delta A$. The second-to-leading and higher
order terms turn out to be finite after summing over $q=t,b$ (for
either relative sign) {\it or} over diagrams (a) and (b) (for only
one of the signs). If we have chosen a correct relative sign, the
divergences in type (b) diagrams should cancel exactly with those
in type (a) diagrams {\it separately} for $q = t, b$. Otherwise
the divergences will add up and the total amplitude will be
logarithmically divergent as the squark masses go to infinity. We
find that the sign in the current note rather than that in
Ref.~\cite{Djouadi:1994js} is favored.

\begin{figure}
\centerline{\epsfysize = 6 in
\epsffile{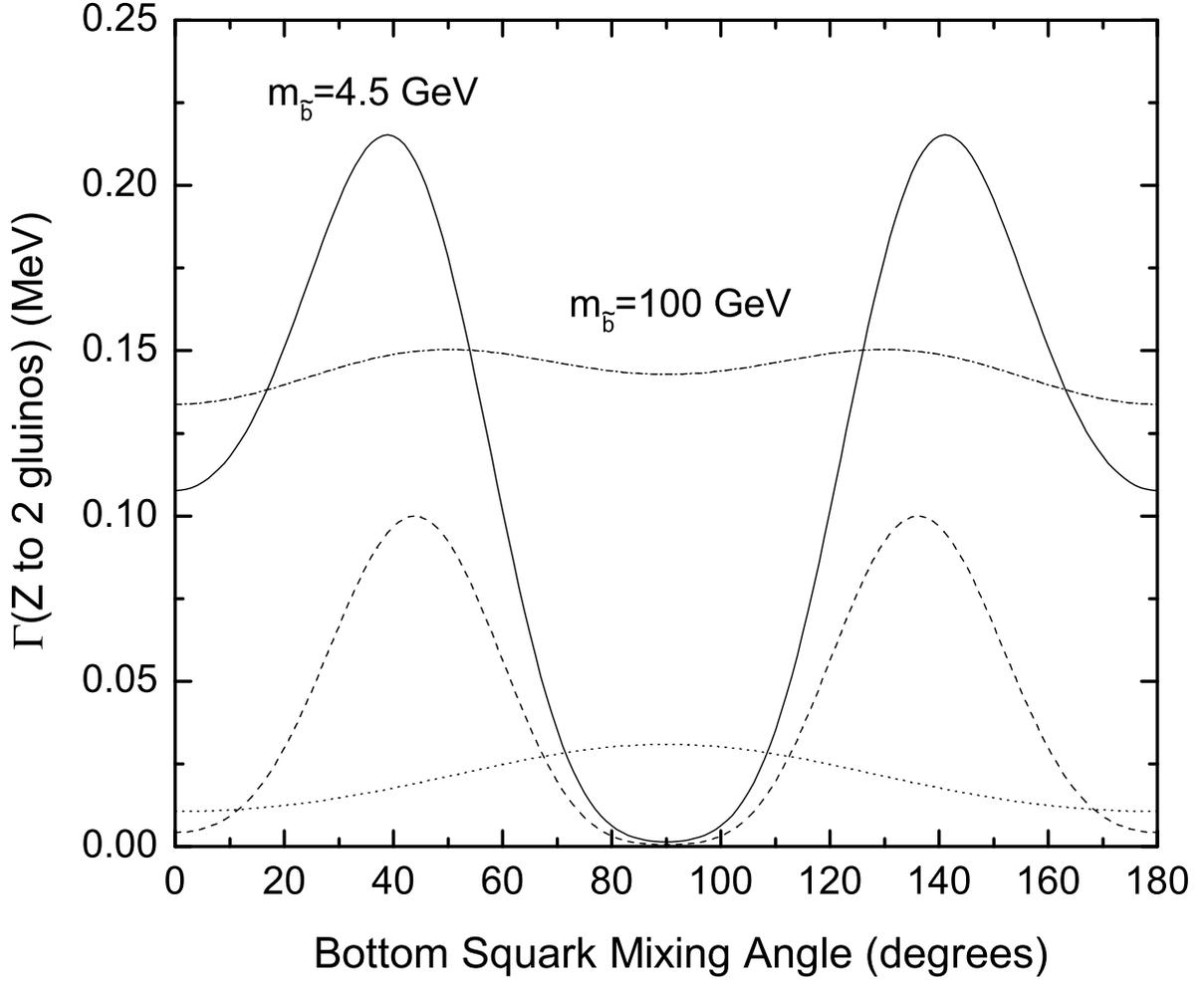}
}
\caption[]{\label{fig:fw}
   $\Gamma(Z \to \sg \sg)$ as a function of the bottom squark mixing
   angle $\theta_{\sb}$. We take $\theta_{\st}=50^\circ$,
   $m_b=4.1$ GeV, $m_{\sb'}=170$ GeV, $m_t=174$ GeV, $m_{\st}=95$
   GeV, $m_{\st'}=300$ GeV, $m_{\sg}=0$. Solid (full width) and
   dashed (unitarity lower bound) curves: $m_{\sb}=4.5$ GeV;
   dash-dotted (full width) and dotted curves (unitarity lower bound):
   $m_{\sb}=100$ GeV. }
\end{figure}

Our numerical analysis not only verifies the above argument but
also shows that it works even if arbitrary $L$-$R$ squark mixing
is allowed and the above constraints on the parameters are
relaxed. Using the same formula in Ref.~\cite{Djouadi:1994js} but
{\it with} the sign flipped \footnote{The formula fails to produce
sensible decay widths in the heavy squark limit if the sign is not
reversed, as argued in the text.}, we find that for $m_{\sg}
\simeq 0$, $m_{\sb} \le {\cal O}(30)$ GeV, $m_{\sb'} = {\cal
O}(150)$ GeV, $m_{\st} = {\cal O}(90)$ GeV, $m_{\st'}= {\cal
O}(300)$ GeV, $\Gamma(Z \to \sg \sg)$ is typically of order 0.1
MeV if $\sb$ is not dominantly left-handed and only of order 0.01
MeV or less if $\sb$ is dominantly left-handed, i.e.,
$\theta_{\sb} \approx 90^\circ$; and that the top squark mixing
angle $\theta_{\st}$ has little effect on the decay width. If all
squark masses are greater than $M_Z/2$, we confirm the statement
in Ref. \cite{Djouadi:1994js} that $\Gamma(Z \to \sg \sg)$ depends
weakly on the details of $L$-$R$ squark mixing and find that it is
of order 0.1 MeV for a wide range of MSSM parameters. Thus
inclusion of a dominantly left-handed light $\sb$ would reduce the
decay width $\Gamma(Z \to \sg \sg)$, as is the case for the
unitarity lower bound plotted in Fig.~\ref{fig:lb}. A relatively
heavy or not very left-handed $\sb$ will not change the decay
width by much. In Fig.~\ref{fig:fw} we plot the full decay widths
as well as the corresponding unitarity lower bounds for
$m_{\sg}=0$ and two sets of squark masses. When all the squark
mass eigenstates are heavier than $M_Z/2$, the full width is about
an order of magnitude larger than the lower bound. When one of the
bottom squarks is light ($\le {\cal O}(30)$ GeV), the shape of the
full width as a function of $\theta_{\sb}$ is similar to the lower
bound and generally only a few times higher than the latter.
Although an expression for $\Gamma(Z \to \sg \sg)$ is not
available when $m_{\sg} \ne 0$, we expect the shape of $\Gamma(Z
\to \sg \sg)$ to be also similar to and only a few times higher
than the unitarity lower bound plotted in Fig.~\ref{fig:lb} (solid
curve) if the gluino has a mass around 15 GeV and the light bottom
squark has a mass around 4.5 GeV. So $\Gamma(Z \to \sg \sg)$
should be of order 0.1 MeV in the light gluino and light bottom
squark scenario.

\section{Implications for gluino searches in $Z$ decays
\label{sec:gs}}

Aside from $Z \to \sg \sg$, there exist three other
gluino-producing $Z$ decays, $Z \to b\bar{\sb}\sg$, $Z \to {\bar
b}\sb\sg$ and $Z \to q{\bar q}\sg\sg$. The first two processes are
$\sim \alpha \alpha_s$ at the tree level and have a combined decay
width of 1.9 -- 5.9 MeV depending on the sign of $\sin
2\theta_{\sb}$ \cite{Cheung:2002na}. The third process is $\sim
\alpha \alpha_s^2$ and its decay width is calculated in a
model-independent way to be 0.75 -- 0.21 MeV for
$m_{\sg}=$12--16 GeV \cite{Cheung:2002rk}. A recent analysis
\cite{Malhotra:2003da} shows that $\Gamma(Z \to b {\bar b}\sg
\sg)$ can be enhanced by 10\% -- 60\% due to additional ``sbottom
splitting'' diagrams. This will raise $\Gamma(Z \to q {\bar q} \sg
\sg)$ by ${\cal O}(0.01)$ MeV. The new SUSY particles do not always
contribute positively to the $Z$ width, however. Cao {\it et al.}
\cite{Cao:2001rz} and S.w. Baek \cite{Baek:2002xf} showed that the
decay width $\Gamma(Z \to b {\bar b})$ can be reduced by 2 -- 8 MeV. By
fine-tuning the parameters in the light gluino
and light bottom squark scenario, all the electroweak measurables
($\Gamma_Z$, $\Gamma_{\rm had}(Z)$, $R_b$, $R_c$) at the $Z$ pole
can be still within the $1 \sigma$ bounds of the experimental
values. Thus, existence of the new particles can only be verified
through direct searches for gluinos or bottom squarks. The light
bottom squark is assumed to be long-lived at the collider scale or
to decay promptly to light hadrons in this scenario. In either
case, it forms a hadronic jet within the detector due to its color
charge. $\sg$ decays exclusively to $b\bar {\sb}$ or ${\bar b}
\sb$ and becomes two hadronic jets. The smallness of $\Gamma(Z \to
\sg \sg)$ implies the insignificance of $Z \to \sg \sg$ in gluino
searches. Searches for signals of $Z \to b \bar {\sb} \sg + {\bar
b}\sb\sg$ and $Z \to q{\bar q}\sg\sg$ will be expected to play a
pivotal role.

\section{Implications for running of $\alpha_s$
\label{sec:ra}}

Both the light gluino $\sg$ and the light bottom squark $\sb$ can
change the $\beta$-function that governs the energy-scale
dependence (``running") of the strong coupling constant
$\alpha_s$. At the two-loop level, $\alpha_s(M_Z)$ can be raised
by $0.014 \pm 0.001$ \cite{Chiang:2002wi} with respect to its
standard model value if extrapolated from the mass scale $m_b$. A
natural question arises: are values of $\alpha_s(M_Z)$ determined
from measurements at different energy scales still in accordance
in the presence of $\sg$ and $\sb$? To answer this question, the
effects of the new SUSY particles on measurements at different
scales must be analyzed. For example, the hadronic width of the
$Z$ is changed in two ways: 1) the interference of the standard
model diagrams and the diagrams with the SUSY particles in loops
will reduce the partial width of $Z \to b {\bar b}$; 2) the
existence of the new decay channels $Z \to \sb \bar {\sb}$, $Z \to
\sg \sg$, $Z \to b \bar {\sb} \sg /{\bar b}\sb\sg$ and $Z \to
q{\bar q}\sg\sg$ will raise the hadronic width. The bottom squark
mixing angle $\theta_{\sb}$ is chosen to be near $23^{\circ}$ or
$157^{\circ}$ so that $\Gamma(Z \to \sb \bar {\sb})$ is suppressed
\footnote{$\Gamma(Z \to \sb \bar {\sb})$ will be greater than 15
MeV if $40^\circ \le \theta_{\sb} \le 140^\circ$.}. $\Gamma(Z \to
\sg \sg)$ is only of order 0.1 MeV at either of these angles. Thus
these two channels combined will change the predicted hadronic
width of the $Z$ by a very small amount compared to the decrease
in $\Gamma(Z \to b {\bar b})$ and the increase in $\Gamma_{\rm
had}(Z)$ due to $Z \to b \bar {\sb} \sg /{\bar b}\sb\sg$ and $Z
\to q{\bar q}\sg\sg$. A better determination of $\Gamma(Z \to b
{\bar b})$, $\Gamma(Z \to b \bar {\sb} \sg /{\bar b}\sb\sg)$ and
$\Gamma(Z \to q{\bar q}\sg\sg)$, or a more precise measurement of
$R_b$ (which will constrain the value of $\Gamma(Z \to b {\bar
b})$ more tightly), is needed for a clear-cut decision in favor of
either the Standard Model or the light gluino and light bottom
squark scenario.

\section{Summary
\label{sec:sum}}

We have calculated the imaginary part of the decay amplitude for
$Z \to \sg \sg$ and used it to analyze the full amplitude and
solve some discrepancies in the literature. We have confirmed the
argument that the decay width vanishes if both quarks and squarks
of a given generation are degenerate in mass (the quarks and
squarks in that generation do not need to have equal mass). We
have found that both divergences in the heavy squark limit
and ultraviolet divergences cancel for each weak isospin
partner, as previously claimed by Refs.~\cite{Campbell:kf, Kane:xp}. We
also favor the relative sign of Refs.~\cite{Berge:2002ev, Campbell:kf}
between diagrams (a) and (b) in Fig. \ref{fig:cut}. Borrowing the formula
for $\Gamma(Z \to \sg \sg)$ from Ref.~\cite{Djouadi:1994js} but with their
relative sign between diagrams (a) and (b) flipped to be consistent with
our calculation, we find that the decay width is of order 0.1 MeV
in the proposed light gluino and light bottom squark scenario.
Compared with other decay processes like $Z \to b \bar {\sb} \sg /{\bar
b}\sb\sg$ and $Z \to q{\bar q}\sg\sg$, $Z \to \sg \sg$ will only play a
moderate role in searches for gluinos and analysis of effects of the SUSY
scenario on $\alpha_s(M_Z)$.

\section*{Acknowledgements}

I would like to thank Jonathan L. Rosner, Cheng-Wei Chiang,
Abdelhak Djouadi and Manuel Drees for very useful discussions and
suggestions. This work was supported in part by the U.\ S.\
Department of Energy through Grant Nos.\ DE-FG02-90ER-40560.

\section*{Appendix: relevant {\cal M}-matrix elements}

We define
\begin{eqnarray*}
A^{(1)} & = &-2 g_s^2\frac{(t^b
t^a)_{ji}}{(p_1-k_1)^2-m_{\sb}^2} \\
A^{(2)} & = &-2 g_s^2\frac{(t^a
t^b)_{ji}}{(p_1-k_2)^2-m_{\sb}^2} \\
{\tilde A}^{(1)} & = &-2 g_s^2\frac{(t^b
t^a)_{ji}}{({\tilde p}_1-k_1)^2-m_b^2} \\
{\tilde A}^{(2)} & = &-2 g_s^2\frac{(t^a
t^b)_{ji}}{({\tilde p}_1-k_2)^2-m_b^2} \\
B_{\pm\pm} & = &\sqrt{(E \pm |{\vk}|)(E \pm |{\vp}|)} \\
{\tilde B}_{\pm\pm} & =& \sqrt{(E \pm |{\vk}|)(E \pm |{\vk}|)} \\
S_{\sb} & = & \sin\theta_{\sb} \\
C_{\sb} & = & \cos\theta_{\sb} \\
\hline\hline
\end{eqnarray*}
${\cal M}$ matrix elements for $Z^{\ua} \to b {\bar b} \to
\sg^{\ua} \sg^{\ua}$:
\begin{eqnarray*}
{\cal M}(Z^{\ua} \to b^{\ua}{\bar b}^{\ua}) & = &
-\frac{g_W}{\sqrt{2}\cos\theta_W}e^{i\phi}\left[
(E-|{\vp}|)g_L^b+(E+|{\vp}|)g_R^b\right]\frac{1+\cos\theta}{2}\delta^{ij} \\
{\cal M}^{(1)}(b^{\ua}{\bar b}^{\ua} \to \sg^{\ua} \sg^{\ua}) & =
& -A^{(1)}
e^{-i\phi}(B_{+-}S_{\sb}-B_{-+}C_{\sb})(B_{+-}S_{\sb}-B_{-+}C_{\sb})
\frac{1+\cos\theta}{2}\\
{\cal M}^{(2)}(b^{\ua}{\bar b}^{\ua} \to \sg^{\ua} \sg^{\ua}) & =
& A^{(2)}
e^{-i\phi}(B_{--}S_{\sb}-B_{++}C_{\sb})(B_{--}S_{\sb}-B_{++}C_{\sb})
\frac{1+\cos\theta}{2}\\
{\cal M}(Z^{\ua} \to b^{\da}{\bar b}^{\da}) & = &
\frac{g_W}{\sqrt{2}\cos\theta_W}e^{i\phi}\left[
(E+|{\vp}|)g_L^b+(E-|{\vp}|)g_R^b\right]\frac{1-\cos\theta}{2}\delta^{ij} \\
{\cal M}^{(1)}(b^{\da}{\bar b}^{\da} \to \sg^{\ua} \sg^{\ua}) & =
& A^{(1)}
e^{-i\phi}(B_{++}S_{\sb}-B_{--}C_{\sb})(B_{++}S_{\sb}-B_{--}C_{\sb})
\frac{1-\cos\theta}{2}\\
{\cal M}^{(2)}(b^{\da}{\bar b}^{\da} \to \sg^{\ua} \sg^{\ua}) & =
& -A^{(2)}
e^{-i\phi}(B_{-+}S_{\sb}-B_{+-}C_{\sb})(B_{-+}S_{\sb}-B_{+-}C_{\sb})
\frac{1-\cos\theta}{2}\\
{\cal M}(Z^{\ua} \to b^{\ua}{\bar b}^{\da}) & = &
-\frac{g_Wm_b}{\sqrt{2}\cos\theta_W}(g_L^b+g_R^b)\frac{\sin\theta}{2}\delta^{ij} \\
{\cal M}^{(1)}(b^{\ua}{\bar b}^{\da} \to \sg^{\ua} \sg^{\ua}) & =
& -A^{(1)}
(B_{+-}S_{\sb}-B_{-+}C_{\sb})(B_{++}S_{\sb}-B_{--}C_{\sb})\frac{\sin\theta}{2}\\
{\cal M}^{(2)}(b^{\ua}{\bar b}^{\da} \to \sg^{\ua} \sg^{\ua}) & =
& A^{(2)}
(B_{--}S_{\sb}-B_{++}C_{\sb})(B_{-+}S_{\sb}-B_{+-}C_{\sb})\frac{\sin\theta}{2}\\
{\cal M}(Z^{\ua} \to b^{\da}{\bar b}^{\ua}) & = &
\frac{g_Wm_b}{\sqrt{2}\cos\theta_W}e^{2i\phi}(g_L^b+g_R^b)
\frac{\sin\theta}{2}\delta^{ij} \\
{\cal M}^{(1)}(b^{\da}{\bar b}^{\ua} \to \sg^{\ua} \sg^{\ua}) & =
& A^{(1)}e^{-2i\phi}
(B_{++}S_{\sb}-B_{--}C_{\sb})(B_{+-}S_{\sb}-B_{-+}C_{\sb})\frac{\sin\theta}{2}\\
{\cal M}^{(2)}(b^{\da}{\bar b}^{\ua} \to \sg^{\ua} \sg^{\ua}) & =
& -A^{(2)}e^{-2i\phi}
(B_{-+}S_{\sb}-B_{+-}C_{\sb})(B_{--}S_{\sb}-B_{++}C_{\sb})\frac{\sin\theta}{2}
\\
\hline\hline
\end{eqnarray*}
${\cal M}$ matrix elements for $Z^{\ua} \to \sb \bar {\sb} \to
\sg^{\ua} \sg^{\ua}$:
\begin{eqnarray*}
{\cal M}(Z^{\ua} \to \sb \bar {\sb}) & = &
\frac{g_W}{\sqrt{2}\cos\theta_W} e^{i\phi}|{\tilde
{\vp}}|\left[g_L^bS^2_{\sb}+ g_R^b
C^2_{\sb}\right]\sin\theta \\
{\cal M}^{(1)}(\sb \bar {\sb} \to \sg^{\ua} \sg^{\ua}) & = &
{\tilde A}^{(1)} e^{-i\phi} |{\tilde {\vp}}|({\tilde
B}_{--}S^2_{\sb}+{\tilde
B}_{++}C^2_{\sb})\sin\theta \\
{\cal M}^{(2)}(\sb \bar {\sb} \to \sg^{\ua} \sg^{\ua}) & = &
-{\tilde A}^{(2)} e^{-i\phi} |{\tilde {\vp}}|({\tilde
B}_{++}S^2_{\sb}+{\tilde
B}_{--}C^2_{\sb})\sin\theta \\
\hline\hline
\end{eqnarray*}

\end{document}